\documentclass[12pt,preprint]{aastex}
\shorttitle{Detection of gamma-rays from clusters}

\begin{document}

\title{A statistical detection of gamma-ray emission from galaxy 
clusters: implications for the gamma-ray background and structure formation.}

\author{Caleb A. Scharf}
\affil{Columbia Astrophysics Laboratory, Columbia University, 550 West 120th 
St, New York, NY 
10027, USA}
\affil{  }
\email{caleb@astro.columbia.edu}
\and
\author{Reshmi Mukherjee}
\affil{Department of Physics and Astronomy, Barnard College, Columbia 
University, New York, NY 10027}
\affil{  }
\email{muk@astro.columbia.edu}

\begin{abstract}The origin of the diffuse extragalactic, high-energy gamma-ray
background (EGRB) filling the Universe remains unknown. The spectrum of this
extragalactic radiation, as measured by the {\sl Energetic Gamma Ray
Experiment Telescope} (EGRET) on-board the Compton Gamma-Ray Observatory
(CGRO), is well-fit by a power law across nearly four decades in energy, from
30 MeV to 100 GeV. It has been estimated that not more than a quarter of the
diffuse gamma-ray background could be due to unresolved point sources. Recent
studies have suggested that much of the diffuse background could originate
from the up-scatter of cosmic microwave background (CMB) photons by
relativistic electrons produced by shock waves in the intergalactic medium
(IGM) during large-scale structure formation.  In this work we search for
evidence of gamma-ray emission associated with galaxy clusters by
cross-correlating high Galactic latitude EGRET data with Abell clusters. Our
results indicate a possible association of emission with clusters at a $\geq
3\sigma$ level. For a subset of the 447 richest ($R\geq 2$) clusters the mean
surface brightness excess is $1.2\times 10^{-6}$ ph cm$^{-2}$ s$^{-1}$
sr$^{-1}$ ($>100$MeV), corresponding to a typical non-thermal bolometric
luminosity of $L_{\gamma}\sim 1\times 10^{44}$ erg s$^{-1}$. Extrapolating
this measurement and assuming no evolution we conservatively estimate that
$\sim 1-10$\% of the EGRB could originate from clusters with $z<1$. For this
cluster population the predicted non-thermal luminosity is in excellent
agreement with our measurement, suggesting that the clusters have experienced
mass accretion within the last $10^9$ yrs. If correct, then future gamma-ray
missions, such as the Gamma-ray Large Area Space Telescope (GLAST) should be
able to directly detect nearby galaxy clusters.

\end{abstract}


\section{Introduction}

During the nine years (1991 - 2000) EGRET was operational on-board the Compton
Gamma-Ray Observatory, it detected a diffuse gamma-ray emission above 30 MeV
filling the Universe. This gamma-ray background is composed of an intense
Galactic component, due to cosmic ray interactions with local interstellar gas
and radiation \citep{hun97}, as well as a diffuse and isotropic extragalactic
component. The extragalactic radiation is well described by a simple power
law with index $-2.1\pm 0.3$ from 30 MeV to 100 GeV \citep{sre98}.  The nature
of this extragalactic gamma-ray background (EGRB) has long been debated.  It
is not clear if the EGRB has a purely diffuse origin or is made up of the
superposition of discrete, unresolved, point sources. The majority of the
identified EGRET sources are active galaxies in the blazar class, and this has
led to the suggestion that the EGRB is produced by an unresolved population of
blazars \citep{pad93}. However, estimates of contributions from unresolved
gamma-ray blazars range from nearly all ~\citep{ste96,ste01} to less than
about a quarter \citep{chi98,muc00}. It seems likely that other sources of the
diffuse extragalactic radiation must exist.

Recent work by ~\citet{loe00} has suggested that some, if not most of the
diffuse gamma-ray background could be a result of large-scale structure
formation in the Universe. Gravitationally-induced shock waves produced
during cluster mergers and large-scale structure formation give rise to
highly relativistic electrons which are responsible for Compton
up-scattering the cosmic microwave background photons to high energy
gamma-rays. The fraction of the shocks' thermal energy transferred to
relativistic electrons could range from 1-10\% ~\citep{wax00}. The
gamma-ray background is then expected to be produced in filaments,
sheets, and extended ($> 1^\circ$) gamma-ray halos associated with newly
formed massive clusters ~\citep{wax00}. In some scenarios, the shock
radii for clusters could be large, with gamma-rays detectable in the form
of 5-10 Mpc-diameter ring-like emission tracing the cluster virialization
shock \citep{kes02}. Other work has suggested that some $0.5-2$\% of the
EGRB could arise solely from the population of clusters ~\citep{col98}.
Recent simulation studies of a $\Lambda$CDM universe suggest that future
high energy gamma-ray experiments such as GLAST or the next generation of
atmospheric Cherenkov telescopes should be able to resolve gamma-rays
from individual clusters in the local Universe ($z\simeq 0.025$,
\citet{kes02}). The association of the EGRB with the large scale
structure of the Universe can then be tested by cross-correlating
gamma-ray maps with known galaxy clusters.

So far, individual galaxy clusters have not been found to be correlated with
discrete gamma-ray sources detected by EGRET, although it has been suggested
that they may be candidates for unidentified EGRET sources \citep{tot00}. In
most cases the predicted gamma-ray fluxes from nearby individual clusters
\citep{dar95,col98} are below the sensitivity limit of EGRET, and indeed,
no excess gamma-ray emission is seen from any individual cluster. For example, a
$2\sigma$ upper limit of $ 4\times 10^{-8}$ photons cm$^{-2}$ s$^{-1}$ at
energies greater than 100 MeV can be derived for Coma, the closest rich cluster
\citep{sre96}. However, in some cases the predicted emissivity from 
individual clusters is in disagreement with the $2\sigma$ EGRET upper limits 
\citep{ens97}.

Previous efforts to constrain the gamma-ray emission of clusters used a
small sample (58) of X-ray luminous clusters \citep{rei99}, and found no
significant signal when the EGRET data centered on the clusters were
co-added. Corrections for Galactic diffuse emission were not made, and
a likelihood analysis of the data found no statistically significant
emission associated with the clusters. \citet{col02} has recently reported
preliminary evidence of the association of unidentified EGRET sources with
galaxy clusters at high latitudes, claiming that these sources also show
strong radio emission. Relativistic particles responsible for the radio
emission are probably also the source of inverse-Compton gamma-rays.  
However, the number of claimed associations is small, and no account was
taken of the known deviation from Poissonian statistics due to
cluster-cluster correlations. As we discuss below, even at high
Galactic latitudes - and even with the removal of the best model of
Galactic emission - residual Galactic signatures remain, which is a known
feature of these models \citep{hun97}. Furthermore, by utilizing the much
larger (albeit more inhomogeneous) optical Abell catalog of clusters and
evaluating emission in radial bins, we increase our search sensitivity by a
factor of 100-1000.

In the present analysis we search for angular correlations of the EGRET
extragalactic diffuse emission with cluster position, using the complete
Abell catalog of clusters (\S2), and radially binning the gamma-ray
emission around the Abell clusters (\S3,4). We then estimate the average
gamma-ray luminosity per cluster and address the question of whether this
is consistent with recent predictions of the gamma-ray energy flux
arising from intergalactic shocks, based on semi-analytic predictions and
hydrodynamical cosmological simulations (\S5,6).

\section{Data Analysis}

We have used archival gamma-ray data from all nine years of EGRET
observations for our present analysis \footnote{as directly available from
the HEASARC archives {\tt http://heasarc.gsfc.nasa.gov/}}.  The maps were
generated by summing data over Phases 1-3 and Cycles 4-9 of the EGRET
observations, using only photons with inclination angles of less than
$30^{\circ}$. EGRET covers an energy range of 30 MeV to over 20 GeV. 
Details of the EGRET instrument and capabilities may be found in
\citet{kan88}. The point spread function (PSF) of EGRET is strongly
energy-dependent, and varies from nearly $6^\circ$ (FWHM) at 100 MeV to
$0.1^\circ$ at $> 1$ GeV \citep{tho93}. EGRET has proved to be a highly
successful gamma-ray experiment, comprehensively surveying the gamma-ray
sky from 1991-2000. The Third EGRET (3EG) Catalog \citep{har99a} lists the
271 point sources detected by EGRET during the first four cycles of its
mission (1991-1995). Since cycle 4, EGRET has been operated in a narrow
field-of-view mode for instrumental reasons \citep{har99b}. Of the 271
point sources in the 3EG catalog, over 60\% are unidentified.

The diffuse gamma-ray emission detected by EGRET consists of two
components, one Galactic, and the other assumed to be extragalactic and
isotropic. A model for the Galactic diffuse gamma-ray emission was
calculated using EGRET data from 1991-1993 \citep{hun97}; a diffuse model
using all nine years of EGRET data is not yet available. Other models of
the diffuse galactic gamma-ray emission, also using subsets of the EGRET
data, have been presented by \citet{str00} and \citet{poh98} (for $>1$ GeV
emission). The model of \citet{hun97} assumes that cosmic rays are coupled
to the interstellar matter density. The gamma-rays are produced mainly by
the interaction of cosmic-ray protons and electrons with the ISM. The
Galactic diffuse component is found to be highly peaked along the Galactic
plane. This model generally describes the EGRET data well on larger scales,
but is less successful for individual regions of the sky. \citet{hun97}
discuss the residual (observed minus best-fit model) intensity in their
analysis of the EGRET galactic diffuse emission. The large scale residual
distributions for energies $> 100$ MeV are less than $\pm 2\times 10^{-5}$
photons s$^{-1}$ cm$^{-2}$ sr$^{-1}$, and are roughly distributed around
zero intensity. For our analysis, we subtract the modeled Galactic
component of the gamma-ray emission from the EGRET gamma-ray intensity
skymaps (see below).

The EGRET data used here have been projected into a $720\times 360$,
($0.5^{\circ}$ pixel size in coordinate units) equatorial grid containing
intensities (photons cm$^{-2}$ s$^{-1}$ sr$^{-1}$) (Figure 1). We also
utilize the maps of raw photon counts in the same projection, in order to
evaluate the Poisson noise associated with each pixel (see below). Our
primary dataset contains photons from all energies $>100$MeV. We have also
analyzed the $>1$GeV dataset (which has a correspondingly smaller PSF),
although the increased photon shot-noise results in no statistically
significant measurements.

Prior to evaluating the cluster-EGRET cross-correlations we typically
perform three sets of clipping or corrections to the data, unless described
otherwise below. First, 2.7$^{\circ}$ radius regions (on the sphere of the
sky) are excised around the 100 {\em identified} EGRET sources to reduce
the overall shot noise level. This corresponds to the 90\% enclosed flux
radius of the full energy weighted ($>100$ MeV, $E^{-2}$ spectrum) EGRET
PSF. Using the 99\% enclosed flux radius of $7.1^{\circ}$ has a negligible
effect on our results. The identified sources consist of 67 blazars and 6
pulsars, as well as 27 sources marginally identified as gamma-ray blazars.

 An additional 171 {\em unidentified}, faint, EGRET sources have been
cataloged.  However, the majority of these are close to the Galactic
Plane within the Galactic Plane cut we describe below (only 41 of all
EGRET sources have $|b|>45^{\circ}$; of those, 22 are identified, 19
unidentified). Excising these additional, unidentified sources has a
negligible effect on our results and we have therefore chosen to remove
only those sources of known origin. As described above, no sources are
identified with known clusters of galaxies.

Second, the region with $|b|<45^{\circ}$ is removed to eliminate the high
intensity Galactic Plane emission. The choice of this cut is determined
by a combination of the desire to minimize the Galactic contamination and
the latitude-dependent incompleteness of the Abell catalog (see below),
and to maximize the number of Abell clusters used in evaluating the
cross-correlation function (thereby maximizing sensitivity and keeping
shot noise to an acceptable level). Variations of $\leq
10^{\circ}$ on this cut have a negligible effect on our results.

Third, the detailed model of the diffuse Galactic emission intensity in the
EGRET data (as described above, \citet{hun97}) is {\em subtracted} from the
masked sky map. Since this model has small residuals, in some instances
(5\%) where the EGRET data is non-zero, the subtraction results in a
negative intensity in a given pixel. The value of the residuals of the
model are scattered about zero (relative to the mean subtracted Galactic
emission), but do seem to exhibit some correlation with galactic latitude,
especially at lower energies, being more negative at lower $b$ and away
from the Galactic center \citep{hun97}. In our analyses we also utilize the
raw EGRET photon count maps to estimate the Poisson, or shot noise, due to
finite photon statistics. Specifically we always construct weighted means
where weights are assigned as $A_i\sqrt{N_i}$, where $N_i$ is the raw
photon number in a given pixel and $A_i$ is the pixel solid angle, which
varies with latitude.  Consequently all pixels with zero photons acquire
zero weight in the calculation of the mean (an unweighted mean, which
simply excludes the zero photon pixels, is found to have a negligible
difference from the weighted mean for our purposes).

For most of the analyses described below we are left
with a total of $\sim 40,000$ usable EGRET pixels ($\sim 10^4$ deg$^{2}$). The
effect of any remaining Galactic emission on our results is discussed below.
The final EGRET dataset is shown in Figure 2. The mean intensity is $2.88 
\times 10^{-5}$ph s$^{-1}$ cm$^{-2}$ sr$^{-1}$.

The catalog of Abell clusters we utilize consists of the original Abell
catalog of 2712 systems, plus the southern ACO extension of 1364 clusters
\citep{abe89}. Figure 3 plots the clusters for comparison to the EGRET
data. We have used the basic Abell parameters of richness ($R$) and
distance class ($D$) in the analysis presented below. Specifically, we
have split the sample between $R\geq 2$ (Figure 3) and $R< 2$ and among
various distance classes. We have also utilized the X-ray selected
Brightest Cluster Survey (BCS) of 206 clusters \citep{ebe98}, combined
with the extended BCS catalog of an additional 108 fainter systems
\citep{ebe00}, although as described below, the larger shot noise of this
smaller X-ray sample prohibits robust measurements.

\subsection{Cross-correlation}

The most rigorous cross-correlation analysis of a surface brightness map
and a collection of points (clusters) would involve a direct summation
over sky cells, smoothing the point distribution by the same kernel (PSF)
as the surface brightness (e.g., as is applied to all-sky X-ray data
\citep{jah91,lah93,miy94}). In the case of the EGRET data, however, we
have chosen a cruder approach of summing, or stacking, gamma-ray emission
in radial annuli centered on all Abell clusters, and then determining the
statistical mean in each angular bin. Ostensibly this does not differ
much from the cell-summation approach, and, unlike cell-summation, we
obtain a direct estimate of the excess emission associated with clusters
by explicitly including the cluster coordinate information. Using the
cell-summation technique would require more extensive modeling to extract
this measure \citep{lah93}. Additionally, given the potentially complex
residual contamination by diffuse, or unresolved, Galactic emission, and
the known selection biases of the Abell catalog, the stacking approach
allows for more intuitive modeling (see \S3.1 below).

Given an equatorial pixel size of $0.5^{\circ}$ for the EGRET data and the
68\% flux enclosure at $1^{\circ}$ radius for the energy weighted $>100$ MeV
PSF, we choose annular bins of width $1^{\circ}$ in our analysis. We note that
the energy dependent PSF can be described as $\theta \leq 5.85^{\circ}
(E_{\gamma}/100 {\rm MeV})^{-0.534}$, where $\theta$ is the energy dependent
radius for 67\% flux enclosure \citep{tho93,esp99}. The area-noise weighted
mean flux excess (hereafter referred to as the mean, see \S2)  above the
global mean (calculated from all unmasked, Galactic-corrected pixels) is then
evaluated by using all unmasked pixels in each radial bin, out to $20^{\circ}$
for each cluster, and then averaging over all clusters used. Pixels are used
if their centers lie within a given annulus, consequently there is some
variation in the number of pixels counted between different cluster centers,
however this variation is negligible on averaging over many objects. We refer
to this angular function as $w_{c\gamma}(\theta)$ or $<\Delta I>=<I-\bar{I}>$
in the results in \S 3 below. In \S 3.1 we assess its significance.

\section{Results}

In Figure 4 we plot our principal results. The curve for all 2469 Abell
clusters with $|b|>45^{\circ}$ peaks at a value of $6.2\times10^{-7}$ ph
s$^{-1}$ cm$^{-2}$ sr$^{-1}$ in the 1$^{\circ}$ bin, and gently declines
with $\theta$. This is generically the behavior expected for a positive
correlation between the two datasets. Upper and lower heavy curves are
for the richest ($R\geq 2$) and poorest ($R<2$) Abell subsets. We note
that, based on the theoretical predictions, more massive cluster systems
are expected to be the sites of higher gamma-ray emission \citep{loe00}.
Consequently, the increased amplitude of the rich cluster $w_{c\gamma}$
relative to that of the poorer clusters goes in the expected sense if
clusters are indeed diffuse gamma-ray sources. The peak amplitude for the
rich cluster subset, in the $1^{\circ}$ bin, is $1.19\times 10^{-6}$ ph
s$^{-1}$ cm$^{-2}$ sr$^{-1}$.
	
We have also investigated the effect of using subsets of the Abell
catalog divided using the distance parameter $D$, which is based on the
brighter galaxy magnitudes in a cluster. The increased noise of using
smaller subsets of clusters makes these measurements less significant.
Using only the more distant ($D> 4$) subset, which is still large, has
little effect on the results presented here, with variations well within
the noise.

In addition, since X-ray-selected clusters are considered to be more
robust in terms of being real, gravitationally relaxed systems, we have
run our cross-correlation using the 304 clusters of the BCS
\citep{ebe98,ebe00}.  However, only 159 systems remain after the Galactic
cuts are applied and the resulting $w_{\gamma c}(\theta)$ is too noisy
to draw any conclusions.

Finally, at energies above 1 GeV the EGRET PSF has a FWHM of only
$0.1^{\circ}$. In principle this could help reduce the effect of Galactic
contamination, and enhance the cross-correlation measure for compact,
resolved, sources associated with clusters. In practice though, the
increased photon shot noise of the $>1$ GeV data result in a statistically
insignificant (although still positive)  $w_{\gamma c}(\theta)$.

\subsection{Monte Carlo Tests}

The strongly energy-dependent EGRET point-spread function \citep{tho93},
in combination with a complex diffuse Galactic emission contribution and
local large-scale structure traced by galaxy clusters (e.g. the
Super-Galactic Plane), indicates that the most direct way to assess the
significance of any measurement is via Monte-Carlo simulations.

Our primary goal is to determine the significance of any positive
correlation between the centroids of local galaxy clusters and the EGRET
intensity map. We therefore perform two sets of Monte Carlo simulations.  
First, we retain the true sky distribution of galaxy clusters but use
"fake" EGRET data. Pixel values are assigned by drawing values randomly
from the real (masked and Galactic model subtracted) data, repeats are
allowed, and photon shot noise is carried with the surface brightness.
Second, we retain the true EGRET data but randomize the galaxy cluster
positions. In both cases we generate 1000 random realizations and perform an
identical angular cross-correlation analysis to that made on the real data.
The population distribution of the angular cross-correlation in each
angular bin is then determined and limits containing 90\%, 95\%, and 99\%
of the realizations, centered about the means, are obtained.

The first approach removes all residual Galactic emission correlations and
all PSF effects, and we consider this to be the true baseline for the case
of uncorrelated noise. The second approach is more conservative, since any
residual Galactic structure is still present and the precise noise
characteristics of the EGRET data are retained.  Our second set of Monte
Carlo simulations generate random position cluster catalogs with the same
number of entries as the real data, although without the measured
cluster-cluster correlation properties. The Abell catalog has a known
angular selection function, due to increasing absorption and increasing
star-galaxy confusion towards the Galactic Plane. This has been fit to a
simple law: $f=10^{\alpha(1-cosec|b|)}$, where $b$ is Galactic latitude and
$\alpha$ has values ranging from $\simeq 0.3$ \citep{bah83} to $\simeq
0.53$ \citep{rom92} depending on the particular subset of clusters modeled.
Here we conservatively apply the larger correction to our simulated
catalogs in order to reproduce the gross features of the Abell cluster
selection.

In Figures 5 and 6 the results from the first set of Monte Carlo
simulations (randomly re-sampled EGRET data) are presented for all
clusters, and for just rich clusters, respectively. In Figures 7 and 8 the
results for the second set of simulations (random Abell catalogs) are
presented in the same format.

For all cases shown in Figures 5, 6, 7, and 8 the measured $w_{\gamma
c}(\theta)$ in the $1^{\circ}$ bin is significant at the $\geq 3\sigma$
level. We note that had clusters exhibited much more significant emission
in the EGRET data then it is likely they would have already been directly
detected. Our result therefore occupies a significance regime inaccessible
to anything but statistical analysis.

\section{Further analysis}

In order to assess more completely the significance of the shape of the
measured $w_{c\gamma}$ we have evaluated the cluster-cluster correlation
($w_{cc}$) for the rich ($R\geq 2$) subset, with identical sky incompleteness
(i.e., the same sky mask as applied to the EGRET data), and using the same
annular binning technique. We also convolve $w_{cc}$ with the EGRET PSF (see
below). $w_{cc}$ and $w_{c\gamma}$ should agree at large angular scales (e.g.
$\sim 10^{\circ}$) (if the correct normalization between cluster sky density
and gamma-ray emissivity is known). At small angular scales $w_{cc}$ should
match $w_{c\gamma}$ only in the case of point-like emission coincident with the
clusters. For moderately extended emission (such as that predicted,
\citet{kes02}), $w_{c\gamma}$ will be suppressed relative to $w_{cc}$ at small
angles - but should still agree at larger scales.

Our measured $w_{cc}$ agrees reasonably well with previous measurements of the
Abell cluster angular auto-correlation \citep{pos86} which follows an
approximate $w_{cc}\propto \theta^{-1}$ relationship. We have not included any
selection bias correction in calculating $w_{cc}$ in order to compare directly
with $w_{c\gamma}$. We also note that neither $w_{cc}$ (before renormalizing)
or $w_{c\gamma}$ drop to zero by $20^{\circ}$.  For this annulus, and 447
clusters in $\sim 10^4$ deg$^{2}$, if the clusters were uniformly distributed
then each data pixel would be sampled several times, and $w_{c\gamma}$ should
equal zero. However, as $w_{cc}$ illustrates, the clusters are not uniformly
distributed. In fact they tend to trace the Super-Galactic Plane which is
contiguous across our entire map regions.

In Figure 9 the results for the richest ($R \geq 2$) subset of clusters are
plotted together with the renormalized cluster-cluster correlation
determined directly from the catalog, before and after smoothing with the
energy weighted EGRET ($>100$ MeV) PSF \citep{esp99}. We have renormalized 
$w_{cc}$ to the $10^{\circ}$ annular bin of $w_{c\gamma}$, where we expect
agreement. We also plot the $>100$ MeV EGRET PSF for illustration. 

 Although the corrected $w_{c\gamma}$ and $w_{cc}$ are broadly similar in
shape, there is clear evidence for a flatter slope in $w_{c\gamma}$. In
particular, at scales less than $\sim 3^{\circ}$ $w_{c\gamma}$ is
significantly lower than $w_{cc}$. At larger scales ($>15^{\circ}$) both
functions are less statistically significant. The results summarized in Figure
9 are consistent with non-pointlike cluster gamma-ray emission extending to
some $\sim 3^{\circ}$. We discuss this result in the context of theoretical
predictions in \S 6.

\section{Constraints on clusters and the EGRB}

The basic theoretical predictions for the IGM origin of the EGRB from
\citet{loe00} are summarized in Eqns 1 \& 2.  Approximately 80-90\% of
the EGRB is predicted to be due to IGM emission. The remaining fraction
can be ascribed to a population of point sources \citep{chi98}. The more
recent work of \citet{kes02} agrees closely with these semi-analytic
predictions, but differs when a numerical cosmological simulation is
utilized (see below). 

Equation 1 is the predicted spectral intensity of the EGRB ($>30$ eV) due
to upscattered CMB photons from relativistic electrons (Lorentz factors
$200<\gamma< 4\times 10^{7}$)  produced in IGM shocks.

\begin{equation}
E^2 \frac{dJ}{dE} = 1.1 \left( \frac{\xi_e}{0.05}\right) \left(\frac{\Omega_b 
h_{70}^{2}}{0.04}\right) \left( \frac{f_{sh}kT}{keV}\right) {\rm keV \; s^{-1} 
cm^{-2} 
sr^{-1}}\;\;\; ;
\end{equation}

where $\xi_e$ is the fraction of shock thermal energy transferred to
relativistic electrons, $\Omega_b$ is the cosmological baryon density
parameter, $h_{70}$ is the Hubble constant in units of $70$ km s$^{-1}$
Mpc$^{-1}$, and $f_{sh}$ is the fraction of baryons shocked to a mass-weighted
temperature $T$ \citep{loe00}. The value of $\xi_e \simeq 0.05$ is obtained
from non-relativistic collisionless shocks in the ISM (SNe) and, as described
in \S1, may actually range from 1-10\% \citep{wax00,kes02}. From cosmological
hydro-dynamical simulations $f_{sh}(kT/keV)\sim 1$ (e.g., \citet{cen99}).

Using a hydro-dynamical cosmological simulation, \citet{kes02} find an
$E^2 \frac{dJ}{dE}$ which is lower than the above semi-analytic,
Press-Schechter-based prediction. The spectrum has a varying slope with
energy and the amplitude varies from 50-160 eV s$^{-1}$ cm$^{-2}$
sr$^{-1}$, in contrast to the 1100 eV in Equation 1. Consequently their
prediction is that only $\leq 15$\% of the EGRB is due to the IGM.
\citet{kes02} attribute this primarily to a lower present-day gas
temperature found in the simulation.

Equation 2 is the predicted non-thermal {\em bolometric} luminosity for
massive clusters of galaxies, again due to CMB upscatter from the accretion
shocks surrounding these systems. For a forming massive cluster
\citep{wax00,kes02};

\begin{equation}
L_{\gamma} = 1.5 \times 10^{45} \left(\frac{\xi_e}{0.05}\right) 
\left(\frac{10^9\;\; yr}{t_{vir}}\right) 
\left(\frac{M_{gas}}{10^{14}M_{\odot}}\right) \left( \frac{kT_{gas}}{5\;\; 
keV}\right) {\rm erg s}^{-1}\;\;\;,
\end{equation}

where $t_{vir}$ is the time taken for gas to cross the cluster
virialization shock or equivalently, a measure of the transience of the
gamma-ray emission; $T_{gas}$ and $M_{gas}$ are the thermal gas
temperature and mass respectively in a cluster.  Both Equation 1 and 2
assume a cosmology with $\Omega_{\Lambda}=0.65$, $\Omega_M=0.35$,
$\Omega_B=0.05$ and $h=0.7$). With a slightly different cosmology (Keshet
et al 2002) where $\Omega_{\Lambda}=0.7$, $\Omega_M=0.3$, $\Omega_B=0.04$
and $h=0.67$, very similar amplitudes, of $1.5$ and $2.2\times 10^{45}$
are determined for Equations 1 \& 2 respectively.

We shall use the excess emission associated with $R\geq 2$ clusters as
measured in the innermost (1$^{\circ}$) bin; $1.19\times 10^{-6}$ ph s$^{-1}$
cm$^{-2}$ sr$^{-1}$ ($>100$ MeV, Figure 4).  This corresponds to the 68\% flux
enclosure of the EGRET PSF, and can therefore be considered as a conservative
choice. Relative to our Galactic model subtracted data this represents a $\sim
4$\% fluctuation above the mean diffuse background, and corresponds to a mean
cluster flux in a $1^{\circ}$ radius aperture of $\sim 1.14 \times 10^{-9}$
photons s$^{-1}$ cm$^{-2}$ ($>100$MeV).

First we estimate the gamma-ray, or non-thermal, luminosity implied for a
rich cluster of galaxies.  Converting our photon count rate into flux
(assuming an $E^{-2}$ spectrum), we obtain $f_{\gamma}=1.87\times
10^{-12}$ erg s$^{-1}$ cm$^{-2}$ ($>100$ MeV). To estimate the {\em
bolometric} gamma-ray flux we then apply a simple correction term. We
assume that the emission spectrum is close to an $E^{-2}$ power law and
that the emissivity runs from $\sim 30$eV to $\sim 1$TeV. The upper limit
of the EGRET sensitivity is $\sim 100$ GeV; thus, correcting for the
missing lower and higher energy flux we obtain $f_{\gamma bolometric}
\equiv f_{30eV-1TeV}=3.55 f_{>100 MeV, EGRET}$. We note that at $\sim 1$
TeV, pair production cascades should effectively cut off the spectrum.
Assuming a mean redshift of $z=0.1$ derived for $R\geq 2$ clusters, and
$h=0.75$ we arrive at an estimated mean luminosity of:

\begin{equation}
\bar{L_{\gamma}}=1 \times 10^{44} {\rm erg s}^{-1}
\end{equation}

For our purposes this is fairly robust. Below we discuss the implications of
our measured $\bar{L_{\gamma}}$.

The local ($z\sim 0$) space density of rich clusters is reasonably well
known e.g., \citet{dep02}, and for $R\geq 2$ is of the order $1\times
10^{-6}$ Mpc$^{-3}$.  Consequently, we can simply derive the local
gamma-ray volume emissivity of rich clusters. Assuming no evolution in
any properties, we can then integrate the expected contribution of rich
clusters to the EGRB from a cosmological volume within $z_{max}$. If
$z_{max}=1$ then we arrive at the estimated surface brightness
(bolometric)  contribution:

\begin{equation}
\bar{I_{\gamma c}}=6.0 \times 10^{-10} {\rm erg s^{-1} cm^{-2} sr^{-1}}
\end{equation}

Integrating the spectrum defined by Equation 1 from 30eV to 1TeV we
arrive at a net predicted bolometric IGM surface brightness of
$I_{IGM}=4.3\times 10^{-8}$ erg s$^{-1}$ cm$^{-2}$ sr$^{-1}$. Therefore,
the measured EGRET flux of rich clusters implies that $\sim 1$\% of the
predicted diffuse IGM may be due to rich clusters with $z<1$.

Clearly this is {\em modulo} many parameters and sources of scatter and
should be treated with appropriate caution. If we consider the
prediction of the Keshet et al (2002) numerical simulations then this
fraction clearly rises to $\sim 10$\% of the IGM emission. Similarly we
have assumed zero evolution in all properties, and that our estimate of
$\bar{L_{\gamma}}$ is unbiased.

For comparison, as described in \S 1, the $2\sigma$ upper limit on
emission from the Coma cluster ($z=0.0231$) is $4\times 10^{-8}$ ph
s$^{-1}$ cm$^{-2}$ ($>100$ MeV). Following the same procedures this
translates to a bolometric luminosity of $L_{\gamma Coma} < 2 \times
10^{44}$ erg s$^{-1}$. However, the space density of Coma-like clusters
($M_{total}\sim 10^{15}M_{\odot}$) is only $3\times 10^{-9}$ Mpc$^{-3}$.
Consequently only some $\sim 0.01-0.1$\% of the EGRB IGM component could
would be expected to arise from the very richest clusters.

Returning to the prediction of Equation 2, what are the other implications
of our measurement? If we assume that we are detecting an actively
accreting rich cluster population (i.e., within $10^9$ yr of shock
formation), then based on our knowledge of Abell clusters the mean thermal
gas temperature of the $R\geq 2$ population is likely to be $\sim 3$ keV
and the typical gas mass is $\sim 10^{13}$M$_{\odot}$ (e.g., \citet{whi97}.
Consequently, evaluating Equation 2 for this population we find that the
predicted $L_{\gamma}$ in in excellent agreement with our measured
$L_{\gamma}$.

\section{Discussion}

We have detected evidence of a  positive correlation of unresolved
gamma-ray emission with clusters of galaxies in the local Universe. The
correlation signal is broadly consistent with emission localized within
$\sim 1^{\circ}$ of clusters, and specifically is more strongly
correlated with optically rich clusters.  There is some evidence that the
gamma-ray emission may be more spatially extended up to $\sim 3^{\circ}$,
or up to $\sim 15$Mpc at the mean redshift of rich clusters in the Abell
catalog. We suggest that the mean non-thermal luminosity associated with
rich clusters is $\bar{L_{\gamma c}}\sim 1\times 10^{44}$ erg s $^{-1}$
(30eV-1TeV) in the local Universe.

Recent theoretical predictions and modeling have suggested that 10-80\% of the
EGRB arises from upscattering of CMB photons by relativistic electrons in the
large-scale gravitational shocks of the IGM, including clusters. If such
models are valid our results suggest that $\sim 1-10$\% of this component of
the EGRB originates from structures associated with rich clusters with
$0<z<1$. Specific predictions for the population of clusters alone have
estimated a contribution to the EGRB of $0.5-2$\%, dominated by systems with
$z\leq0.2$ \citep{col98}.

Several caveats are worth noting. First, diffuse Galactic gamma-ray emission
is very hard to remove, even with the current best model for its distribution.
We believe we have accounted for the residual effects of this emission by
assessing significance via our Monte Carlo simulations, but it may still
introduce systematic errors to our results. Second, we cannot, with these
data, prove that the excess emission associated with clusters is not due to
some population of discrete sources, themselves positively correlated with
clusters, although there is no a priori reason to believe this to be the case.
Third, estimating the cluster luminosity by measuring the intensity in the
innermost 1$^{\circ}$ bin is likely to be an underestimate, given both the
width of the EGRET PSF and the evidence for more extended, correlated,
emission. Finally, several potential sources of random error exist serving to
increase the uncertainty in quoted numbers - which, therefore, should be
considered as broad estimates.

It should also be noted that another possible mechanism for producing
non-thermal emission in clusters is synchrotron emission in radio galaxies
contained in the clusters, that produce relativistic electrons which then
up-scatter to produce the high energy gamma-rays (e.g. \citet{gio93,pet01}).
In this case the particles are accelerated in the nucleii of radio galaxies,
and this could account for the observed gamma-ray luminosities. In this case,
one would expect to see a stronger correlation signal with rich clusters
containing radio galaxies. \citet{led95} find that some 79\% of $R=2$ Abell
clusters have evidence of radio emitting member galaxies (in a statistical
subsample of 393 clusters of all richnesses).  However, more sophisticated
analysis is required owing to the small number of radio surveyed clusters and
we will investigate this in an extension to the present work.

As discussed by \citet{col98}, \citet{loe00}, \citet{wax00}, and
\citet{kes02}, the appearance of non-thermal emission in IGM/cluster
structures is transient, with a timescale of $\sim 10^{9}$yrs. Therefore
the emission disappears almost entirely as soon as there is no strong
shocking of gas. If we consider our sample of rich Abell clusters then we
know from other wavelengths that the majority of these low-$z$ systems
are probably in a state closely approximating hydrostatic equilibrium
(e.g., the presence of X-ray cooling flows, the nearly isothermal gas,
etc.). We can consider two extreme states: either only some,
non-equilibrium clusters are gamma-ray bright, or all clusters are close
to equilibrium and we are simply seeing the fading emission of an earlier
epoch of accretion activity. In the first case we can exploit the fact
that a system with an EGRET flux of at least $\sim 5 \times 10^{-8}$ ph
s$^{-1}$ cm$^{-2}$ ($>100$MeV) would be directly detectable in the EGRET
map. Given our mean flux of $\sim 1.1\times 10^{-9}$ (\S5) and a total of
447 rich clusters, we would obtain an equivalent detection if only $\sim
10$ rich clusters had a flux of $5 \times 10^{-8}$ and the rest were
gamma-ray dark. This would then provide a lower limit to the number of
actively accreting clusters in the local Universe, a 2\% fraction by
number.

Alternatively, if all clusters are assumed to be close to equilibrium with
a mean temperature of $kT\simeq 3$keV, a mean gas mass $\simeq
10^{13}$M$_{\odot}$, and, $\xi_e=0.05$ then they are almost exactly as
gamma-ray luminous as predicted by Equation 2. This implies that in fact
all clusters should have been actively accreting recently, certainly within
$z<0.3-0.4$, and possibly at the present time ($z=0$).

Even in the low-density simulation of \citet{kes02} it appears to be the
case that $z=0$ clusters can have active gamma-ray emission.  
Semi-analytic predictions for low density cosmologies ($\Omega_m=0.3$,
$\Omega_{\Lambda}=0.7$) also suggest that even at $z\sim 0$, for massive
clusters ($10^{15}$ M$_{\odot}$) several $10^{13}$M$_{\odot}$ in baryons
should accrete per $10^{9}$yr \citep{lac93}. The shock regions may form
some 5-10Mpc from the cluster core, creating a gamma-ray `ring' of emission
\citep{kes02}. The suggestion of a rather more extended emission pattern
from our cross-correlation (Figure 9) supports this scenario. In this case
our estimated $\bar{L}_{\gamma c}$ can be considered a good measure of the
typical active non-thermal emission for rich clusters.  This would then
imply that the efficiency of transfer of energy from the shocks to
relativistic electrons is similar to the value of $5$\% inferred from
non-relativistic shocks in the ISM.

The simulations of \citet{kes02} predict that for $\xi_e\geq 0.03$,
future high resolution gamma-ray telescopes with threshold sensitivities
$>10^{-10}$ for energies $> 10$ GeV should be able to resolve some
gamma-ray halos associated with large scale structures. The prospect of
direct detection of gamma-ray sources with emission attributed to
intergalactic shocks with GLAST, VERITAS, HESS, MAGIC, or other
atmospheric Cherenkov telescopes is an exciting one.  Such gamma-ray
halos could be a new source class of high energy sources waiting to be
discovered. Their existence would allow an entirely new, and direct,
probe of structure formation processes, leading to an improved
understanding of inter- and intra-cluster gas dynamics, magnetic fields,
and energy partitioning. While extrapolations from EGRET data and
simulations for future instruments predict at least a dozen or more
detectable sources \citep{kes02}, direct determination of gamma-ray
sources due to shocks can be used as an independent calibration for
$\xi_e$. In fact, if the value of $\xi_e$ were lower than the inferred
5\% (see discussion above), this would have an impact on the number of
sources resolvable by future telescopes. \citet{kes02} predict that the
number drops to $\sim 1$ for $\xi_e\sim 0.2$. GLAST will have the
spectral and spatial resolution to confirm whether galaxy clusters can be 
directly detected in gamma-rays.

\acknowledgements

We thank D. Helfand for originally suggesting this work and for his invaluable
comments, and F. Paerels, S. Digel, and E. Waxman for discussions which have
helped improve this paper. C. S. acknowledges support from NASA grant
NAG5-6035, R. M. acknowledges support from NSF grant PHY-9983836.

This research has made use of data obtained from the High Energy Astrophysics
Science Archive Research Center (HEASARC), provided by NASA's Goddard Space
Flight Center.

\begin{figure} 
\caption{The $>100$MeV EGRET all-sky intensity map in
rectangular equatorial projection.  The Galactic Plane is clearly seen as the
brightest region in the sky; some individual bright sources are also 
evident.
Galactic South is towards the lower center of this image.}
\plotone{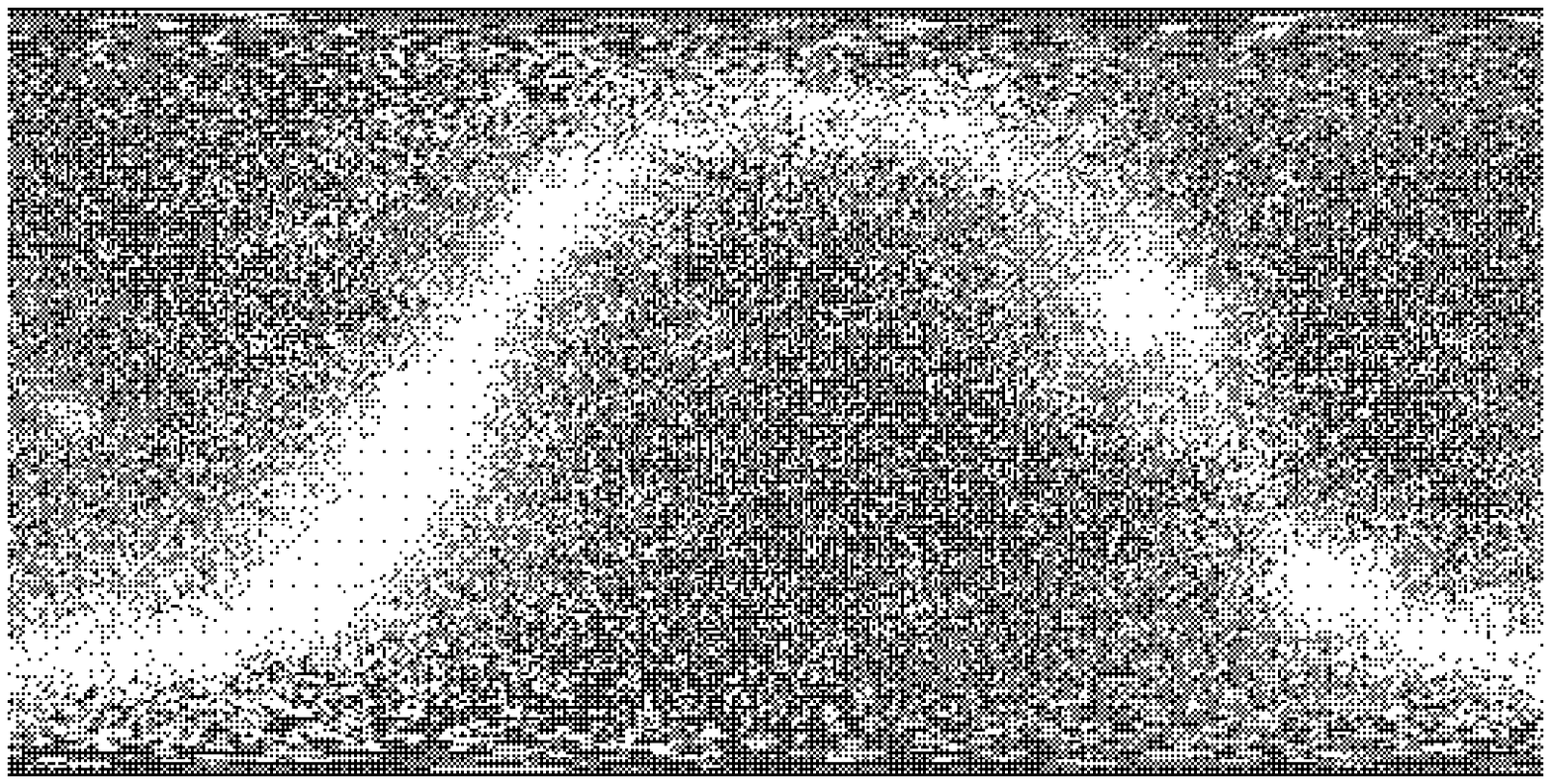} 
\end{figure}

\begin{figure} 
\caption{EGRET data as in Figure 1, but with all mask cuts and
the diffuse Galactic model intensity subtraction applied. Two of the excised
bright sources (center and left in figure) clearly have remaining extended flux
in this map.  However, this flux has negligible impact on our analysis since by
area it is small.}
\plotone{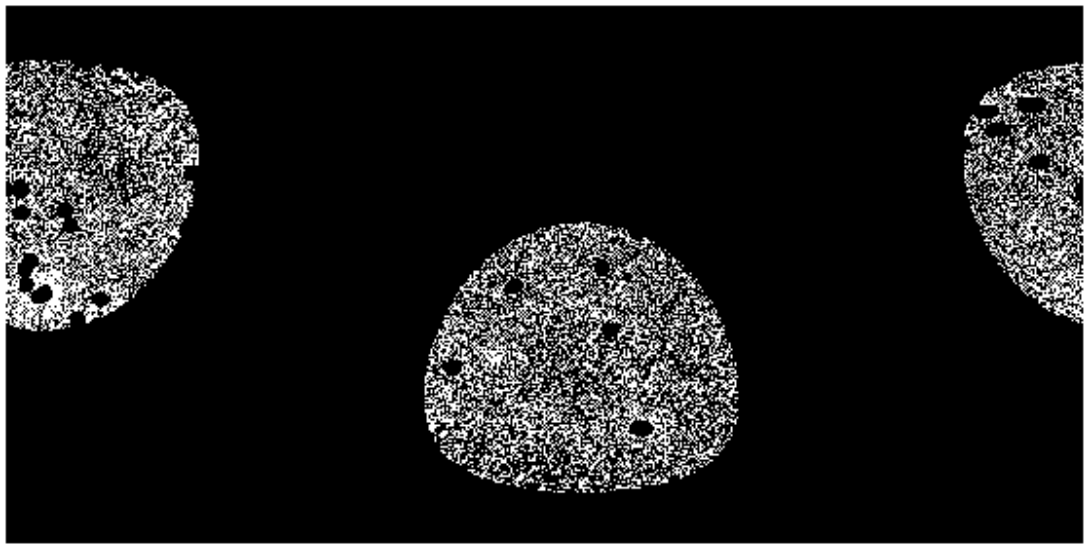}
\end{figure}

\begin{figure}
\caption{Upper panel: All 4076 Abell clusters are plotted in rectangular 
equatorial coordinates as for the EGRET data. The strong selection bias 
away from the
Galactic Plane is reflected in the sparsity of clusters in those regions. Lower 
panel: The 447 richest ($R\geq 2$) Abell clusters are plotted.}
\plotone{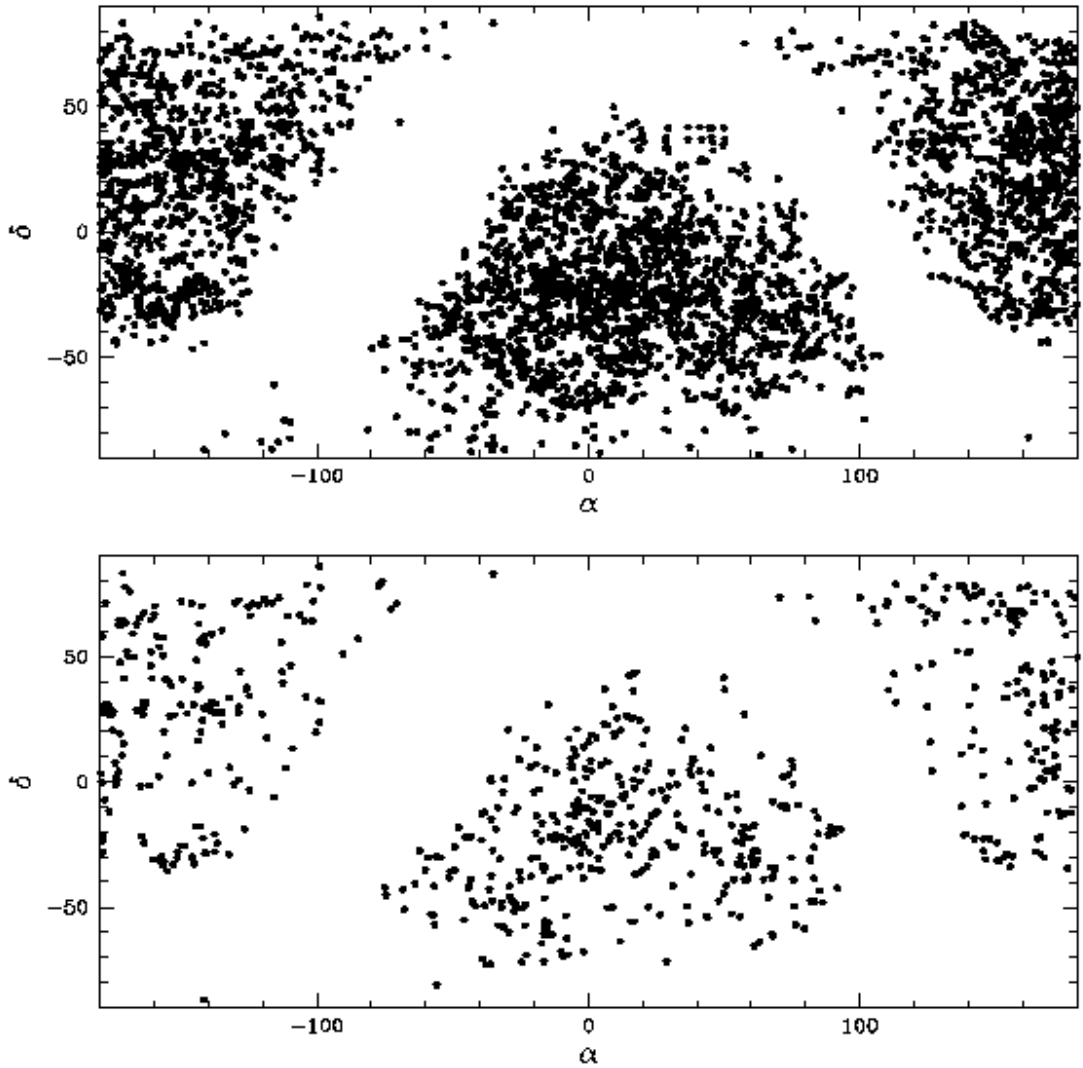}
\end{figure}

\begin{figure}
\caption{The mean excess $>100$MeV emission ($<\Delta I>=<I-\bar{I}>$), 
or 
cross-correlation $w_{c\gamma}(\theta)$,
in radial bins of width 1$^{\circ}$, surrounding Abell clusters. 
The lightest solid 
curve is the result for all Abell clusters within the usable EGRET data 
regions, a total of 2469 systems. The uppermost curve uses only the 447
richest ($R\geq 2$) clusters overlapping the EGRET regions, while the
 lowermost curve is the corresponding measurement for all clusters with 
$R<2$. The horizontal dot-dashed line indicates the zero-point.}
\plotone{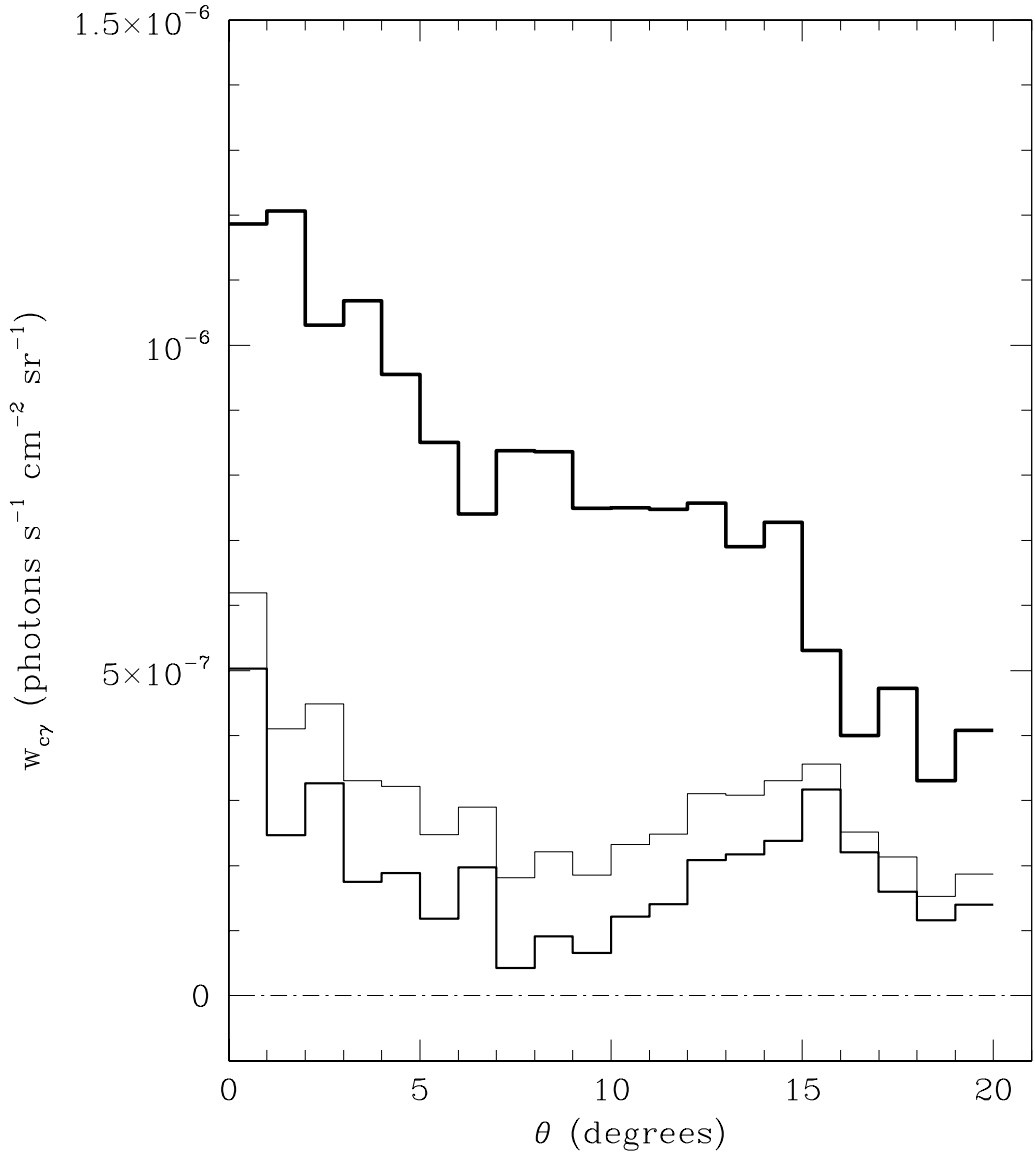}
\end{figure}

\begin{figure}
\epsscale{0.7}
\caption{Monte Carlo results for 200 realizations with scrambled EGRET data 
and the full Abell catalog (2469 clusters). The distribution of the mean 
excess emission ($w_{c\gamma}$) is plotted in grayscale. Lightest gray 
indicates the limits within which 90\% of the realization $w_{c\gamma}$'s 
lay, 
medium gray indicates the 95\% population, and the darkest gray the 99\% 
population. As expected, the mean over all realizations is zero at all 
$\theta$. The heavy solid line is the measured Abell-EGRET $w_{c\gamma}$}
\plotone{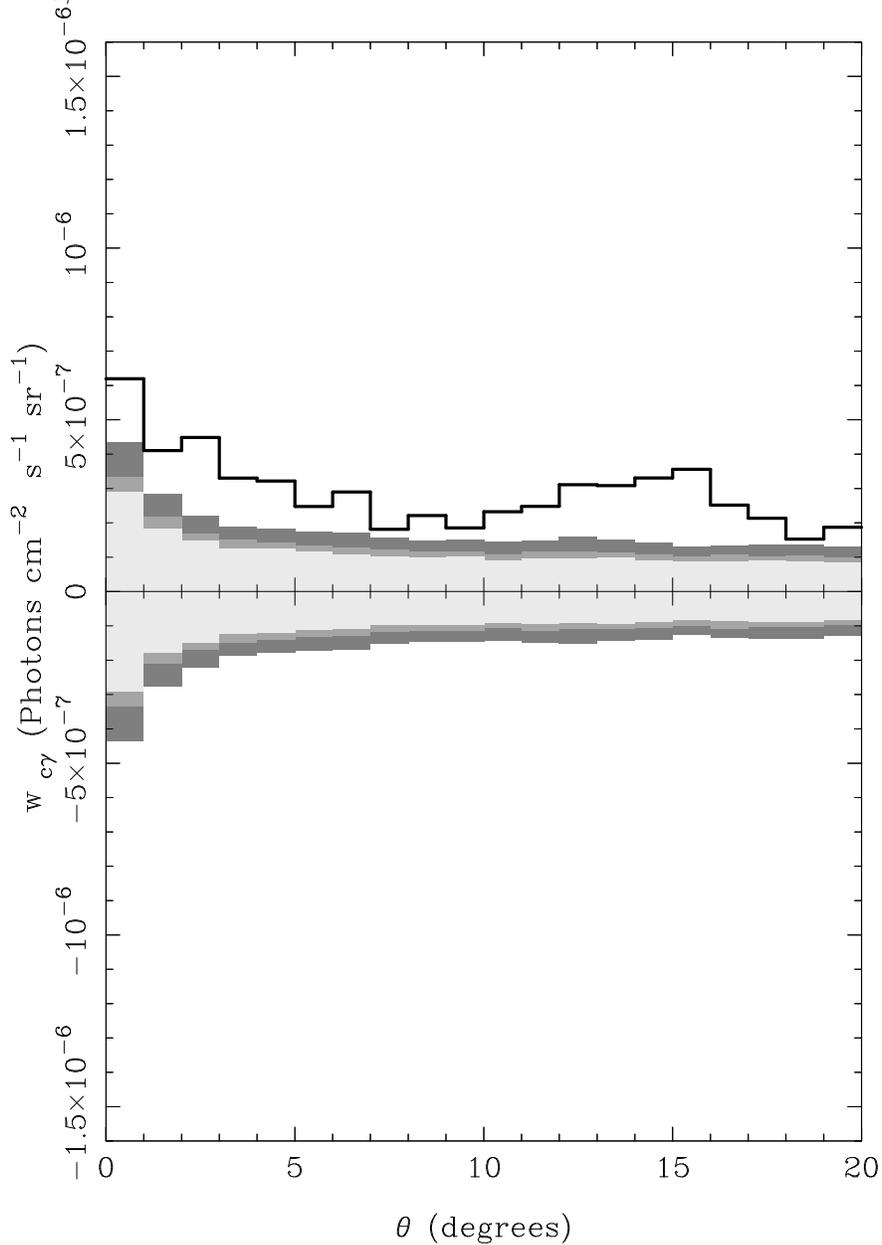}
\end{figure}

\begin{figure}
\caption{As for Figure 5, but now for the 447 richest, $R\geq 2$ 
clusters only.}

\plotone{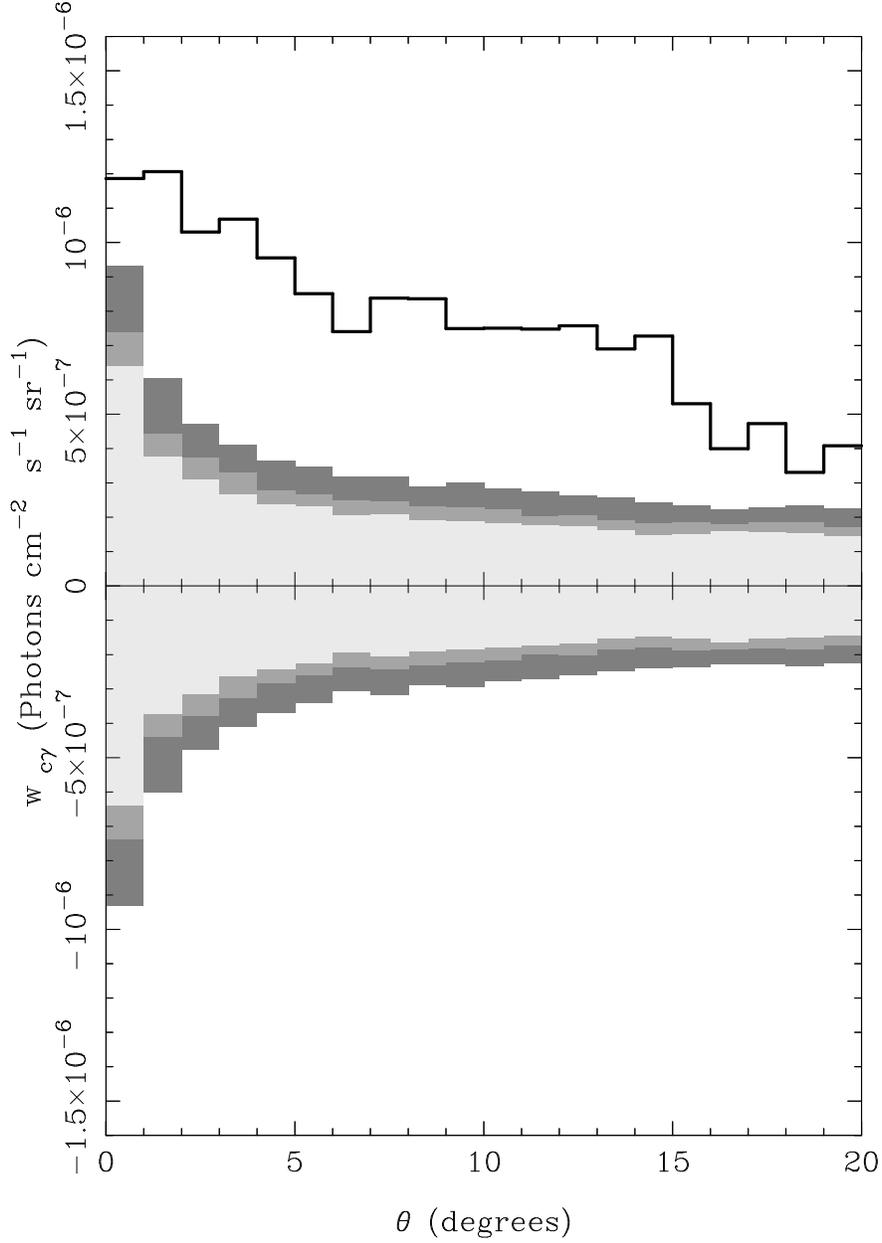}
\end{figure}

\begin{figure}
\caption{Monte Carlo results for 200 realizations with random Abell catalogs 
and the full, unscrambled, EGRET data. As in Figures 5 and 6 the gray scale 
denotes the 90\%, 95\% and 99\% population limits, the heavy solid line is
the measured $w_{c\gamma}$ for all 2469 Abell clusters.}

\plotone{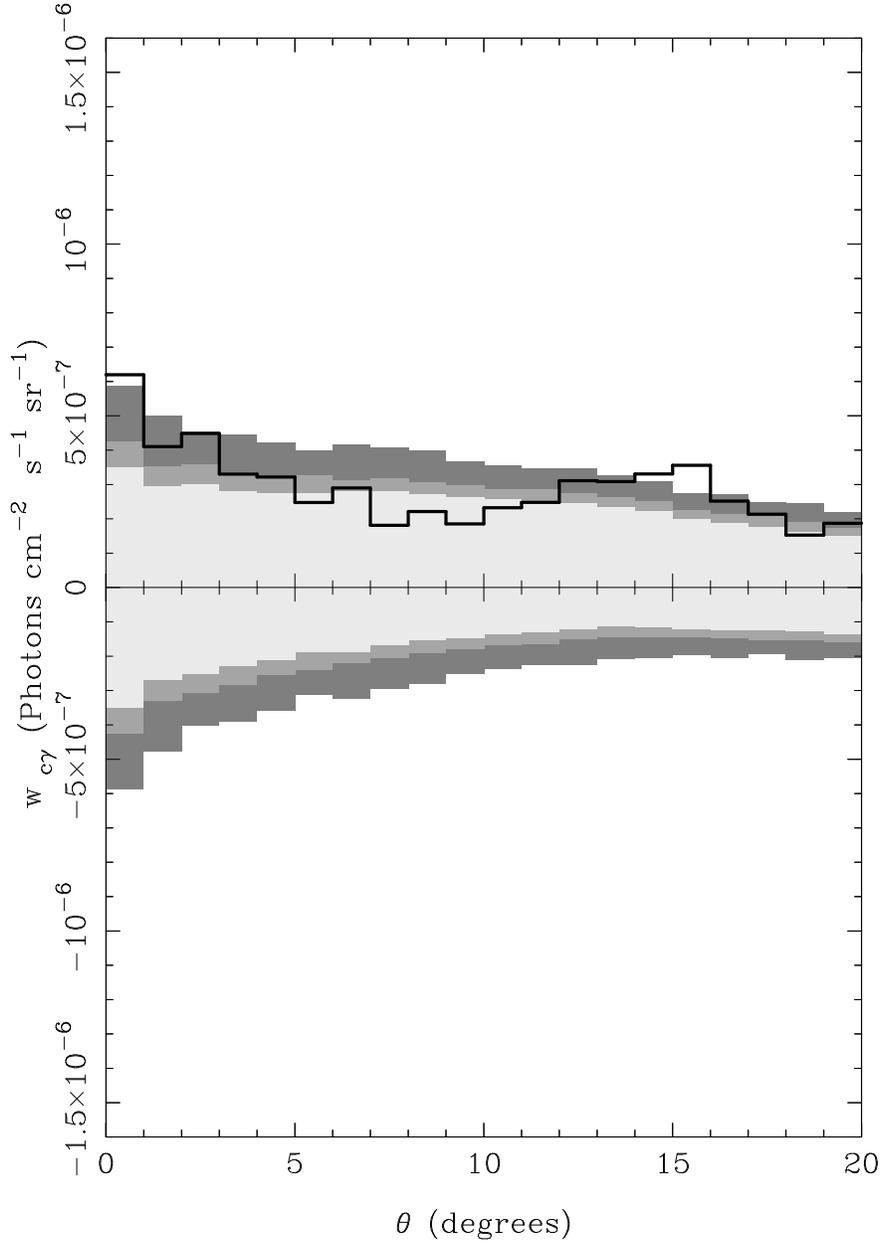}
\end{figure}

\begin{figure}
\caption{As for Figure 7, but now for the 447 richest, $R\geq 2$
clusters only.}

\plotone{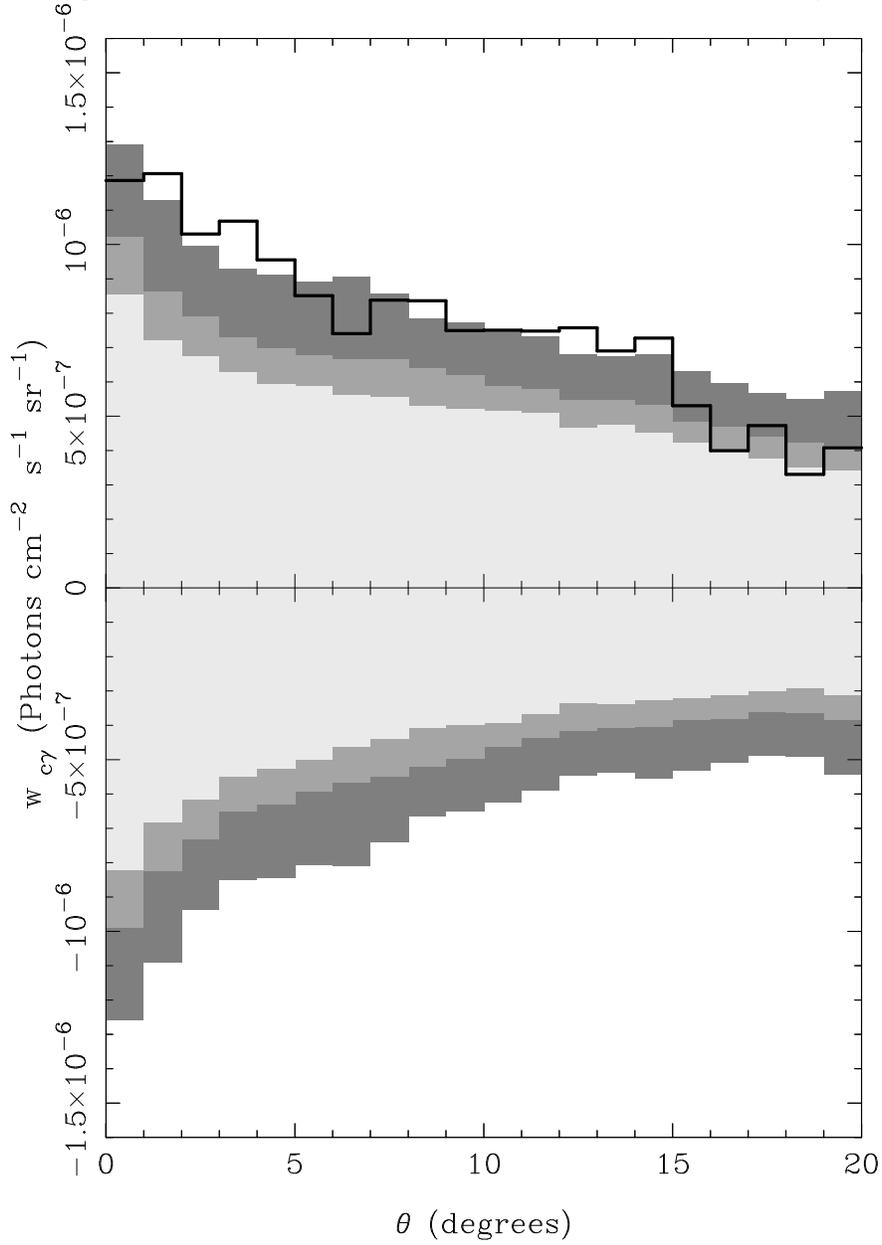}
\end{figure}

\begin{figure} 
\epsscale{1.0} 
\caption{$w_{c\gamma}$ (or $<\Delta I>$) is
plotted for the subset of 447 richest ($R \geq 2$)  clusters (heavy solid
line). The measured $w_{cc}$ for this subset of clusters, re-normalized to
match the 10 degree bin of $w_{c\gamma}$ (see text), is shown (dotted line),
together with its convolution with the energy-weighted $>100$MeV EGRET PSF
(thin solid line). The $>100$ MeV EGRET PSF, with arbitrary normalization is
also plotted (dashed line).} 
\plotone{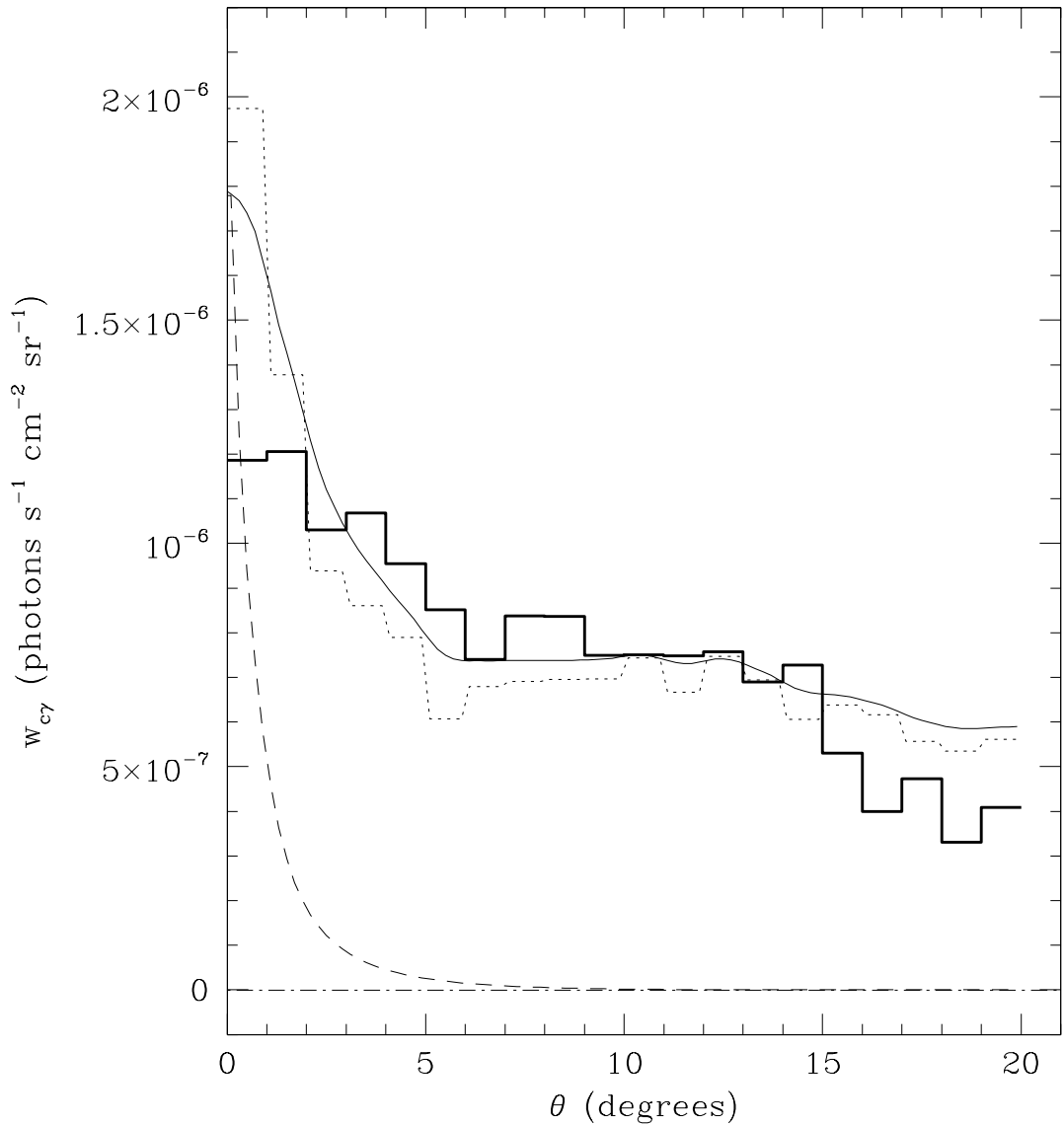} \end{figure}

\end{document}